\documentclass[twocolumn]{aastex6}
\usepackage{graphicx}
\usepackage{amsmath}
\usepackage{epstopdf}
\usepackage{natbib}
\received{August 11, 2017}
\shorttitle{Magnetic pumping as a source of heating}
\shortauthors{Lichko et al.}

\begin{document}

%\preprint{}

\title{Magnetic pumping as a source of particle heating and power-law distributions in the solar wind}


%\thanks{}
%\affiliation{Department of Physics, UW-Madison, WI, 53706, USA}
%\affiliation{Los Alamos National Laboratory, Los Alamos, New Mexico, 87545, USA}
%\affiliation{Department of Atmospheric, Oceanic, and Space Sciences, University of Michigan, Ann Arbor, MI, 48109, USA}
%\affiliation{Harvard-Smithsonian Center for Astrophysics, Cambridge, MA, 02138, USA}

\author{E. Lichko}
\affiliation{Department of Physics, UW-Madison, WI, 53706, USA}
\author{J. Egedal}
\affiliation{Department of Physics, UW-Madison, WI, 53706, USA}
\author{W. Daughton}
\affiliation{Los Alamos National Laboratory, Los Alamos, New Mexico, 87545, USA}
\author{J. Kasper}
\affiliation{University of Michigan, Ann Arbor, MI, 48109, USA}

%\keywords{text}
%\affiliation{Harvard-Smithsonian Center for Astrophysics, Cambridge, MA, 02138, USA}

%\author{W. Daughton}
%\affiliation{Los Alamos National Laboratory, Los Alamos, New Mexico
%87545, USA}

\newcommand{\scr}[1]{_{\mbox{\protect\scriptsize #1}} }

\begin{abstract}

Based on the rate of expansion of the solar wind, the plasma should cool rapidly as a function of distance to the Sun. Observations show this is not the case. In this work, a magnetic pumping model is developed as a possible explanation for the heating and the generation of power-law distribution functions observed in the solar wind plasma. Most previous studies in this area focus on the role that the dissipation of turbulent energy on microscopic kinetic scales plays in the overall heating of the plasma. However, with magnetic pumping particles are energized by the largest scale turbulent fluctuations, thus bypassing the energy cascade. In contrast to other models, we include the pressure anisotropy term, providing a channel for the large scale fluctuations to heat the plasma directly. In this work a complete set of coupled differential equations describing the evolution, and energization, of the distribution function are derived, as well as an approximate closed form solution. Numerical simulations using the VPIC kinetic code are applied to verify the model's analytical predictions. The results of the model for realistic solar wind scenario are computed, where thermal streaming of particles are important for generating a phase shift between the magnetic perturbations and the pressure anisotropy. In turn, averaged over a pump cycle, the phase shift permits mechanical work to be converted directly to heat in the plasma. The results of this scenario show that magnetic pumping may account for a significant portion of the solar wind energization. 

% the anisotropic structures to decay more quickly than one would expect from just pitch-angle scattering

\end{abstract}

%\pacs{}

%\maketitle

\section{Introduction} \label{sec:intro}

Spacecraft measurements of the solar wind indicate the presence of an anomalous source of heating throughout the heliosphere. The radial temperature distribution observed not only exceeds the values expected given the rapid expansion of the solar wind within the Parker spiral \citep{richardson:2003, kasper:2002, cranmer:2009, stverak:2009}, but also the heating that would be expected from steady-state hydrodynamic models \citep{parker:1965, hartle:1968, durney:1972}. Much of the work on this topic in the past decade has focused on how energy is injected as turbulent fluctuations at large spatial scales, and then propagates through the turbulent cascade and is absorbed by the plasma at the smaller dissipation scale \citep{chandran:2010, told:2015, howes:2008, howes:2011, bruno:2013}. While there exists direct evidence for the turbulent cascade operating in the solar wind \citep{ sahraoui:2009}, there are significant details of the solar wind heating problem that are not readily explained by either the energy cascade or traditional stochastic acceleration mechanisms. One such detail in particular is the observed power-law in the solar wind's particle distributions, $f \propto v^{\gamma}$ where $\gamma = -5$, which is present throughout the solar wind \citep{fisk:2006,fisk:2012}. 

In this Letter we propose a heating mechanism based on magnetic pumping, a process by which a series of either periodic or random magnetic perturbations heats a plasma. While it in no way precludes energy transfer through the turbulent cascade, this model allows energy transfer to the particles directly from plasma perturbations on the largest scales, bypassing the turbulent cascade. It also accounts for many of the observations seen in the solar wind, such as the power-law distributions in velocity space, that have been observed in the solar wind. 

The most important physical aspect of the model, and the aspect in which it differs most from other models for heating in the solar wind, is the role of pressure anisotropy in relation to the fluctuating magnetic field. To elucidate this mechanism, one may consider a simple magnetic flux tube. In the absence of heat fluxes and other isotropizing effects, contractions of the tube lead to pressure anisotropy in phase with the density and magnetic perturbations \citep{chew:1956}, such that there is no net energization. If, however, an isotropizing effect is present there will be a net energization of the plasma, directly related to a finite phase difference between pressure anisotropy and the plasma compressions, yielding positive work by the term $P_{\perp} \nabla_{\perp} \cdot  \bf{v}$ when averaged over a pump cycle. This result is consistent with recent results from \citet{yang:2017a, yang:2017b} that investigate the importance of the $(P \cdot \nabla) u$ term in the collisionless turbulent cascade.

For the one-dimensional flux tube described above, the spatially isotropizing effects can be introduced through pitch-angle scattering. However, in a two-dimensional scenario, where the plasma perturbations include temporal as well as spatial variations, pressure anisotropy is reduced by parallel thermal heat fluxes. In this scenario if an area of increased anisotropy is created locally, it decays due to the thermal streaming of particles in and out of the region. Thus, anisotropy in $f(\bf{v})$ with spatial variations of $l_{\scr{pert}}$ decays at the rate $\nu_{\scr{eff}}\sim l_{\scr{pert}}/v$. Below, when applying the model to the solar wind for realistic $dB/B(\omega)$ spectra, we find that $\nu_{\scr{eff}}\sim l_{\scr{pert}}/v$ dominates the ion isotropization, permitting significant ion energization and the associated formation of power law distributions. 

\section{Comparison to other models} \label{sec:PhysDescrModel}  

Many models address the heating and formation of power law distributions in the solar wind \citep{fisk:2006,fisk:2012,chandran:2010}, but do not adequately explain the observed levels of energization \citep{lynn:2013}. In constrast to that body of work, the present analysis considers heating channeled to the plasma by pressure anisotropy. We note that the mathematical framework is similar to that applied by \citet{drake:2013} for Fermi acceleration during magnetic island coalescence, where magnetic or density perturbations yield velocity changes $\Delta v \propto v$. The analysis then brings about a velocity diffusion equation of the form
\begin{equation*}
\frac{d f}{d t}=K(\tau_{\scr{pump}},\nu_{\scr{eff}})\frac{1}{v^2}\frac{d}{d v} \left(v^2D\frac{df}{d v}\right)
\end{equation*}
where $D\propto (\Delta v)^2/\tau_{\scr{pump}} \propto v^2/\tau_{\scr{pump}}$ is the diffusion coefficient and $K$ is a function with a maximum at $1/\tau_{\scr{pump}}\simeq \nu_{\scr{eff}}$, providing a measure of the level of anisotropy displayed by $f$.  

The model should not be confused with transit-time damping \citep{stix:1992}. In our model the heating efficiency is proportional to $(dB/B)^2$ and is derived by considering a standing wave, for which wave-particle resonances are unimportant. With the standing wave being comprised of two oppositely propagating compressional waves, the heating can be interpreted as a nonlinear interaction between these two waves. While such interactions are fundamental to the development of plasma turbulence \citep{chandran:2009}, the described heating channel is typically neglected with the assumption of isotropic velocity distributions. 

Furthermore, in a standard gyrokinetic treatment the distribution function is assumed to consist of a background Maxwellian distribution, constant in time, as well as a time-dependent first order perturbation. While we do start with a Maxwellian distribution, in our model we allow the ``background'' distribution to vary in time, which is critical to realizing the power-law formation. 

It is worth noting that magnetic pumping has been investigated in the context of fusion research where Coulomb collisions provide pitch-angle scattering, but also relax the heated distributions into Maxwellians \citep{laroussi:1989, borovsky:1990, berger:1958}. However, in the solar wind particles may pitch-angle scatter collisionlessly off of instabilities and fluctuations \citep{verscharen:2016, bale:2009}. This occurs with negligible energy diffusion \citep{kulsrud:1968} such that Boltzmann's H-theorem does not apply \citep{boltzmann:1872}, and Maxwellian distributed particles are generally not observed. We show below that in this limit of collisionless scattering, magnetic pumping yields power-law solutions with spectral indices consistent with values observed \textit{in situ} in the solar wind.

\section{Kinetic Simulations} \label{sec:KineticSimulations} 

\begin{figure}[ht!]
\plotone{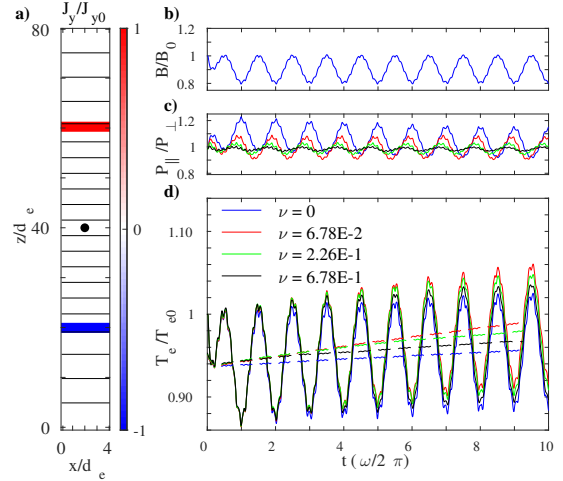}
\vspace{0.0cm}
\caption{\textbf{(a)} A representation of the simulation domain. The colored regions are the points where the external current is applied. The width of the applied current region is increased 2.5x for visualization purposes. The black circle represents the z-coordinate where the measurements in (b-d) were obtained.\textbf{(b)} Plot of the magnetic field taken at $z=40d_e$ \textbf{(c)} Plot of the pressure anisotropy taken at $z=40d_e$ for $\nu/\omega_{pump}$ = 0, 6.78E-2, 2.26E-1,and 6.78E-1. (See legend in (d)) \textbf{(d)} Plot of temperature taken at $z=40d_e$. Temperature measurements were obtained by taking \(T_e=tr(P)/(3 n_e)\). The solid lines are the absolute results and the dotted lines are the average taken over one period.} \label{fig:JyModBTe_AllNu}
\end{figure}

Our initial set-up is a one-dimensional flux tube, as shown in Fig.~\ref{fig:JyModBTe_AllNu}(a). The domain is doubly periodic and an external, sinusoidally driven current is applied along two infinite current sheets each located halfway between the mid-line and the top and bottom edge of the simulation space. The oppositely directed current sheets cause flux tube expansions and contractions as the current oscillates (as show in Fig.~\ref{fig:JyModBTe_AllNu}(b)). The background distribution is given by a Maxwellian with uniform temperature, \(T_e=T_i=T_0\) with the mass ratio given by \(m_i/m_e = 100\). The simulations used a non-relativistic thermal speed $v_{the}/c=0.0707$ and $\omega_{pe}/\Omega_{e0}=1$. Spatial scales are normalized by $d_e$, and our pumping frequency, $\omega_{pump}$, is normalized by $\omega_{pe}$. In the simulations presented below we use $\omega = \omega_{pump}=0.1~\omega_{pe}$, where $\omega_{pump}$ is referred to hereafter as $\omega$. Our density and magnetic field fluctuations are normalized by the initial density, $n_0=1$, and background magnetic field, $\mathbf{B_0}=B_0\hat{x}=1$, respectively. For the purposes of this initial work, the scattering frequency, $\nu$, is velocity-independent and is implemented using the Takizuka and Abe Monte Carlo method employed to calculate Coulomb collisions in VPIC \citep{takizuka:1977, daughton:2009}. Only electron-ion collisions are included, so the energy diffusion is minimal given the mass ratio. The 1D flux tube simulations were carried out in VPIC for a variety of scattering frequencies. From Fig.\ref{fig:JyModBTe_AllNu}(d) it is clear that magnetic pumping is increasing the temperature of the plasma. Furthermore, both the energization and the phase difference between pressure anisotropy and magnetic field that gives rise to the plasma energization show a dependence on scattering frequency, $\nu$. 

\section{Analytical Model Derivation} \label{sec:AnalyticDerivation}

We next proceed to derive an analytic model to explain the energization and demonstrate that it matches the results from the kinetic simulations. To that end, we consider a periodic flux tube with length $l$ and radius $r$. The plasma within the flux tube is assumed to remain uniform while $r$ and $l$ change slowly in time such that the magnetic moment $\mu=m{v_\perp^2}/(2B)$ of the particles is conserved. Furthermore, given the periodic boundary conditions, the action integral $J=\oint v_{\parallel} dl$ is also an adiabatic invariant provided that the length of the tube is not changed significantly during a typical particle transit. 

For this system, our first aim is to obtain a reduced  drift kinetic equation, $df/dt=0$, where 
\begin{equation}
\frac{d}{dt}=\frac{\partial }{\partial t} + \frac{\partial v_\perp^2}{\partial t}\frac{\partial}{\partial v_\perp^2} + \frac{\partial v_{\parallel}}{\partial t}\frac{\partial}{\partial v_{\parallel}},
\end{equation}
is the total time derivative along the particle trajectories. Note that given the assumption of a uniform plasma the convective spatial derivative term (${\bf v} \cdot\nabla$) vanishes.
 
Assuming $\mu\propto {v_\perp^2}/B$, $J\propto l v_{\parallel}$, and $N=\pi r^2nl$, we use the conservation of magnetic moment, action, and particle number to rewrite the above time derivative. For the simple geometry considered, the drift kinetic equation $df/dt=0$ simplifies to 

\begin{equation}
\label{eq:dfdt}
\frac{\partial f}{\partial t}+\frac{\dot{B}}{B}{v_\perp^2}\frac{\partial f}{\partial {v_\perp}^2}
+\left(\frac{\dot{n}}{n}-\frac{\dot{B}}{B}\right){v_{\parallel}}\frac{\partial f}{\partial v_{\parallel}}=0\quad.
\end{equation}

In this limit without scattering, we note that evolution equations for ${p_\parallel}$ and ${p_\perp}$ are readily derived by calculating the ${v_{\parallel}}^2$ and ${v_\perp^2}$ moments of of Eq.~\ref{eq:dfdt}, which yield the CGL double adiabatic scaling laws
\begin{equation}
\label{eq:CGL}
{p_\parallel}\propto n^3/B^2 \,\,,\quad {p_\perp} \propto nB \quad.
\end{equation}

In the following, we explore changes induced in $f$ due to uniform perturbations of the flux tube in conjunction with steady pitch-angle diffusion limiting the development of pressure anisotropy. Thus, similar to \citet{drake:2013} we generalize our kinetic equation to include additional physical effects
\begin{equation}
\label{eq:dfdtCf}
\frac{df}{dt}= \nu {\cal L} f - c_1\,f+ c_2 \,  f_{ext}\quad,
\end{equation}
where ${\cal L}={\partial}/{\partial \zeta}(1-\zeta^2){\partial}/{\partial \zeta}$ is the Lorentz scattering operator, $\zeta={v_{\parallel}}/v$ is the cosine of the pitch-angle, and $\nu$ is a typical frequency for the  scattering processes. The constants $c_1$ and $c_2$ specify the rate of plasma losses and rate of incoming (external $f_{ext}$) plasma, respectively.

For the analysis below it is convenient to change variables from $(v_{\parallel},{v_\perp^2})$ to $(v,\zeta)$. Eq.~\ref{eq:dfdtCf} then takes the form: 
\begin{eqnarray}
\label{eq:dfdtC}
\frac{\partial f}{\partial t}&+&R\left(P_2(\zeta)v\frac{\partial f}{\partial v}\,+\,
\frac{3}{2}\zeta(1-\zeta^2)\frac{\partial f}{ \partial \zeta}\right)\,+\,\frac{\dot{n}}{3n}v\frac{\partial f}{\partial v}
\nonumber\\[2ex]
&=& \quad \nu {\cal L} f - c_1\,f+ c_2 \,  f_{ext}\quad,
\end{eqnarray}
where $P_2(\zeta)$ is the second order Legendre Polynomial and $R = (2/3) {\dot n}/{n} - {\dot B}/{B}$. We note that ${d}/{dt} [\log ({p_{\parallel}}/{p_{\perp}} )] = {d}/{dt} [\log ({n^2}/{B^3} )] = 3R$, showing that $R^3$ is proportional to the rate at which the pressure anisotropy builds in the CGL system (the system with $\nu=c_1=c_2=0$).

To evaluate the efficiency by which the plasma is energized in the above framework, we next consider periodic perturbations for the magnetic field and density. An approximate solution to Eq.~\ref{eq:dfdtC} can be obtained by expanding $f$ in a series of Legendre polynomials $f(v, \zeta, t) =\sum_j P_j(\zeta)f_j(v,t)$ where $P_j$ is the $j$th order Legendre polynomial. The approach provides a set of coupled differential equations, which we solve numerically, and a first order approximation to the results. These two solutions will then be compared to the results of the kinetic simulations.

We still consider the uniform and periodic flux tube but now with imposed sinusoidal temporal variations in density and magnetic field:
\begin{eqnarray}
\frac{\dot{n}}{n}&=&\frac{\delta n}{n} i\omega e^{i\omega t} \,\,,\quad
\frac{\dot{B}}{B} \,=\, \frac{\delta B}{B} i\omega e^{i(\omega t+{\phi_B})}\,\,, \nonumber  \\[2ex]
R&=&i\omega\delta R\,e^{i\omega t}\,\,,\,\,\delta R=\left|\frac{2}{3}\frac{\delta n}{n}-\frac{\delta B}{B}e^{i{\phi_B}}\right|\quad.
\end{eqnarray}

Our aim is again to obtain a solution to Eq.~\ref{eq:dfdtC}. Given the periodic variations of the drive, contrary to the analysis in \citet{drake:2013} we do not need to impose an ordering involving $\nu$, but only require that $\delta R \ll 1$. By inserting the above expansion in pitch-angle into Eq.~\ref{eq:dfdtC} and integrating over $\int_{-1}^{1} P_n(\zeta) d \zeta$, we can obtain a set of coupled differential equations for an arbitrary order of Legendre polynomial. These equations can be solved numerically to arbitrary precision. In the following comparisons, all numerical solutions were taken to second order, truncating at the $f_2$ equation, as further terms resulted in negligible improvements in accuracy. 

Additionally we obtain an approximate solution by assuming that each $f_n$ is comprised of a slowly varying component and a rapidly varying component, denoted hereafter as: 
\begin{equation}
f_n = f_n^s(v,t) + \tilde{f_n}(v) e^{i \omega t} \quad.
\end{equation}

Inserting this approximation into Eq.~\ref{eq:dfdtC} and collecting terms proportional to $P_2(\zeta) e^{i \omega t}$ we obtain the relation: 
\begin{equation}
\label{eq:tft}
{\tilde{f}_2}=K\,v\frac{\partial f_0^s}{\partial v}\quad,\quad K= -\frac{\omega\, {\delta}R\, (\omega + i 6 \nu)}{\omega^2+36\nu^2} \quad.
\end{equation}

Eq.~\ref{eq:tft} shows how the $P_2$-perturbation of the distribution develops and will, for finite $\nu$, be offset in phase from the drive oscillation in $R$, by the angle $\theta = \arctan(6 \nu / \omega )$. This phase shift is important because when solving for $f_0$ we obtain non-vanishing time averages from the terms involving ${\tilde{f}_2}$. Since ${\cal E} = {3}/{2} \int  v^2 f d^3 v = 6 \pi \int f_0 v^4 dv$, these non-vanishing terms become the source of the energization. Using Eq.~\ref{eq:tft}, an equation is obtained for the slowly varying ``background'' distribution: 

\begin{equation}
\label{eq:dfdtHeat2}
\frac{\partial f_0^s}{\partial t} - \frac{3}{5}\,\frac{\nu\,(\omega\, {\delta}R)^2}{\omega^2+36\nu^2} \frac{1}{v^2} \frac{\partial}{\partial v} v^4 \frac{\partial f_0^s}{\partial v} = - c_1 f_0 + c_2 f_{ext} \quad.
\end{equation}

For velocities sufficiently large that the cold source is negligible, the solutions to Eq.~\ref{eq:dfdtHeat2} then take the form:

\begin{equation}
\label{eq:formFSoln}
f_0^s \propto v^\gamma,\quad \gamma= -\frac{3}{2} - \sqrt{\frac{9}{4} + \frac{c_1}{G}}, \quad G = \frac{3}{5}\,\frac{\nu\,(\omega\, {\delta}R)^2}{\omega^2+36\nu^2} \quad.
\end{equation}

In the limit of no net losses (\textit{i.e.} $c_1 = 0$), the exponent, $\gamma$, approaches $-3$. Thus, the heating mechanism is more than adequate to account for the observations of $f \propto v^{-5}$ distributions typically observed in the solar wind \citep{fisk:2006, fisk:2012}. %\newline

From Eq.~\ref{eq:dfdtHeat2} we can directly obtain the heating range

\begin{eqnarray}
\label{eq:dfdtHeat3}
\frac{\partial {\cal E}}{\partial t}
 &=&\frac{3}{2} G
\int_0^{\infty} v^2 \frac{1}{v^2}\frac{\partial}{\partial v}\left(v^4\frac{\partial f_0}{\partial v}\right)  4\pi v^2 dv = 10 G {\cal E},~~
\end{eqnarray}
which gives us

\begin{equation}
{\cal E} = e^{\frac{6 \nu (\omega \delta R)^2}{\omega^2 + 36 \nu^2} t} \quad.
\end{equation}

When we combine the above expression with the analytic solution for the $\nu = 0$ case, ${\cal E}_0(t)$, which is essentially sinusoidal, to obtain a solution for the energy of the system for arbitrary $\nu$ that agrees with the simulation:  

\begin{equation}
\label{eq:energyEqn}
{\cal E}(t) = {\cal E}_0(t) e^{\frac{6 \nu (\omega \delta R)^2}{\omega^2 + 36 \nu^2} t} \quad.
\end{equation}

\section{Discussion and application to the solar wind} \label{sec:Discussion}

\begin{figure}[ht!]
\plotone{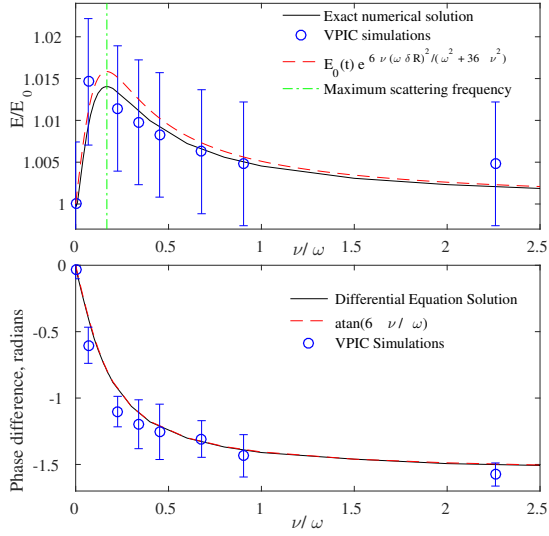}
%\vspace{0.3cm}
\caption{\textbf{(a)} Comparison of fractional energy increase as a function of scattering frequency for the results of the kinetic simulations and the two analytical methods. Error bars are determined using the standard deviation of individual particle energies. \textbf{(b)} Comparison of phase difference between $P_{\parallel}/P_{\perp}$ and B as a function of scattering frequency. Phase difference is normalized relative to the $\nu = 0$ phase difference. Error bars are determined using the standard deviation of the phase difference calculated for each period in simulation.}
\label{fig:PhaseDifferencesWTheoreticalPredictions}
\end{figure}

Given the above results, it is now possible to compare the predictions of our analytic model with the results from the kinetic simulations. The relationship between the relative energy evolution (${\cal E}/{\cal E}(t=0)$) and the scattering frequency is shown in Fig.~\ref{fig:PhaseDifferencesWTheoreticalPredictions}(a). Again there is good agreement between the VPIC simulations and both the exact numerical results of our analytic model as well as the results from the first order approximation. Based on the form of Eq.~\ref{eq:energyEqn}, the scattering frequency that will maximize the energization is obtained, as shown in Fig.~\ref{fig:PhaseDifferencesWTheoreticalPredictions}(a). The analytic solutions and VPIC results all peak at this most efficient frequency, further lending credence to the agreement between the models. Similarly, for the phase difference between $P_{\parallel}/P_{\perp}$ and $B$ there is a good agreement between the two analytic solutions and the VPIC simulations, as in Fig.~\ref{fig:PhaseDifferencesWTheoreticalPredictions}(b).

%Physically we would expect there to be a most efficient scattering frequency. If $\nu = 0$ , then the effects of magnetic pumping do not play a role in the simulations at all, and we have the situation described in the beginning of this paper, with no net heating. As $\nu \rightarrow \infty$, the system isotropizes too quickly, and again there is no net heating. 

\begin{figure}[ht]
\plotone{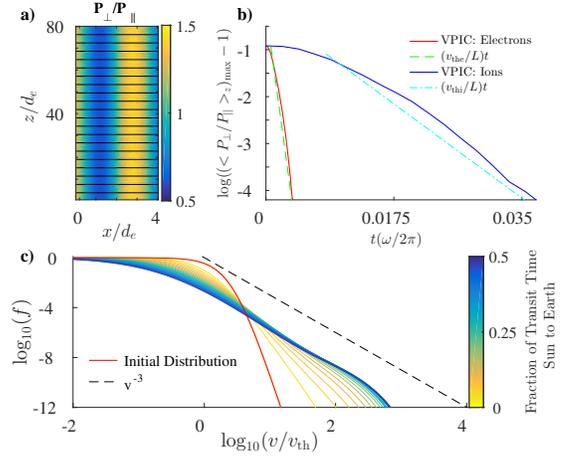}
\vspace{0.0cm}
\caption{ \textbf{(a)} VPIC initial set-up for the phase-mixing rate test. Black lines are field lines of the background magnetic field. \textbf{(b)} Results from the VPIC simulations of phase-mixing.\textbf{(a)} Numerical solution for the distribution function after many oscillations. Note that the slope approaches the value commonly observed in the solar wind, $\gamma=-3$. In this plot $v_{\scr{th}}=90~\mathrm{km/s}$.} \label{fig:PwrLaw}
\end{figure}

For the solar wind, scattering is infrequent and the main isotropizing effect for the pressure anisotropy is thermal streaming, for which we estimate $\nu_{\scr{eff}}\sim l_{\scr{pert}}/v$. To verify this estimate we set up a VPIC simulation using the same domain as the 1D simulations described above, but with no magnetic fluctuations. We initialized the domain with a spatially dependent anisotropy, as shown in Fig.~\ref{fig:PwrLaw}(a), and observed the decay rate. The results of this for both electrons and ions are shown in Fig.~\ref{fig:PwrLaw}(b). It is clear that the decay rate matches our expectations of $\nu_{\scr{eff}}\sim l_{\scr{pert}}/v_{\scr{th}}$. A similar phenomena has been used in other models, such as in \citet{hammett:1990} where it is used to form a closure to the MHD equations. The isotropization caused by thermal streaming is much larger than that induced by pitch-angle scattering off waves and Coulomb collisions.

However, to estimate $\nu_{\scr{eff}}$ properly for the solar wind we need to take into account the spatial anisotropy, \textit{i.e.} that $k_{\perp}>>k_{\parallel}$ \citep{chen:2016}. Since this particle streaming effect is only in the parallel direction, we need to take into account only the part of the $l_{\scr{pert}}$ that is parallel to the magnetic field. As such, where $k=1/l_{\scr{pert}}$, the correct form of the isotropizing effect is $\nu_{\scr{eff}}\sim v k_{\parallel}\sim v/l_{\scr{pert}}(k_{\parallel}/k)\sim v/l_{\scr{pert}}(k_{\parallel}/k_{\perp})$. 

\begin{figure}[ht]
%\vspace{0.5cm}
%\begin{center}
	%\includegraphics[width=8.0cm]{1Da2DHeating.eps} 
%\end{center}
\plotone{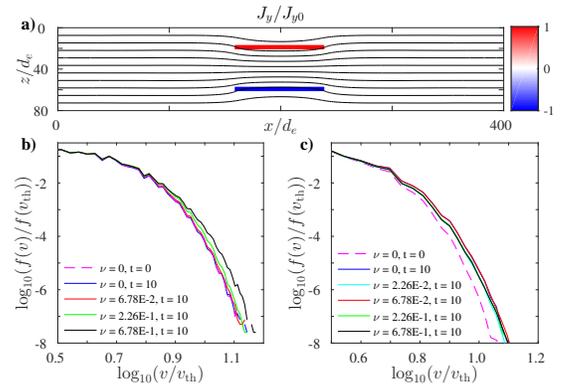}
\caption{\textbf{(a)} Simulation domain for the 2D simulations. Notice that the current sheets only stretch across one-fifth of the domain. \textbf{(b)} Log-log plot of the distribution function for the 1D simulation. As seen in the earlier plots, the level of heating is dependent on the scattering frequency, $\nu$. \textbf{(c)} Log-log plot of the distribution function for the 2D simulation. Note that the heating is now independent of scattering frequency.} 
\label{fig:ModBAnisotropy_AllNu}
\end{figure}

The results of the model, using parameters relevant to the solar wind ions, are shown in Fig.~\ref{fig:PwrLaw}(c). To generate this plot a selection of frequencies were randomly generated and their corresponding $dB/B(\omega)$ were taken from the spectra in \citet{leamon:1998}. The distribution function was then evolved using the set of coupled differential equations described above in a closed system ($c_1=c_2=0$). After every cycle a new randomly generated frequency and corresponding $dB/B(\omega)$ was chosen, so that every cycle the plasma would experience a new frequency, consistent with the observation that fluctuations in MHD turbulence decohere after a single cycle. From \citet{chen:2016} Fig.~7 we used  $k_{\parallel}/k_{\perp}\sim8$ when generating Fig.~\ref{fig:PwrLaw}(c). Using a solar wind speed of 800~km/s we obtain an estimate of the total transit time from the Sun to the Earth. We also assume that only $10\%$ of the perturbations that the plasma will experience will be compressional. As shown in Fig.~\ref{fig:PwrLaw}(c), we obtain a power-law distribution out to two orders of magnitude in velocity after the plasma experiences only half the cycles it would be expected to experience on its way from the Sun to the Earth.

As an additional test of the model we implement a kinetic simulation of a 2D pump geometry. As shown in Fig.~4, the set-up is the same as in Fig.~\ref{fig:JyModBTe_AllNu}(a), but the domain has been extended in the $x$ direction, and the current sheets providing the oscillating magnetic perturbations only cover a portion of the simulation domain. In this case, the heating is no longer dependent on the scattering frequency, $\nu$, as it was in the 1D simulations shown in Fig.~\ref{fig:ModBAnisotropy_AllNu}(b). Thermal streaming of electrons out of the pumping region acts as an effective scattering process, leading to an increased amount of heating.

\section{Conclusions} \label{sec:Conclusions}

In summary, we have explored magnetic pumping as a possible heating mechanism of the solar wind. Energy associated with large scale magnetic and density fluctuations heats the plasma directly. The energization is related to the phase of the pressure anisotropy being shifted from the turbulent drive through the inclusion of collisionless pitch-angle mixing. This phase difference between the pressure anisotropy and the driving magnetic perturbations serves as the source of the heating, and an important point of distinction between this model and previous models of energization in the solar wind. Bypassing the turbulent cascade, the model provides an efficient scheme for dissipating energy directly with the largest scale fluctuations and generating power-law particle distributions.

\acknowledgments
This work was supported by NASA Grant No. NNX15AJ73G. Contributions from E.L. were conducted with Government support under and awarded by the DoD, Air Force Office of Scientific Research, National Defense Science and Engineering Graduate (NDSEG) Fellowship, 32CFR168a. Simulations were performed with LANL institutional computing resources.

\end{document}